\documentclass[superscriptaddress,amsmath,amssymb,aps,prl,reprint,nofootinbib]{revtex4-2}

\usepackage{graphicx}
\usepackage{dcolumn}
\usepackage{bm}
\usepackage{hyperref}
\usepackage{notes2bib}
\usepackage{siunitx}
\usepackage{bbold}
\usepackage[usenames,dvipsnames]{xcolor}

\newcommand{\Rxx}{\ensuremath R_\mathrm{xx}}
\newcommand{\Rxy}{\ensuremath R_\mathrm{xy}}
\newcommand{\Vtg}{\ensuremath V_\mathrm{tg}}
\newcommand{\Vbg}{\ensuremath V_\mathrm{bg}}

\definecolor{PR}{rgb}{0.57, 0.36, 0.51}
\definecolor{todo}{rgb}{1, 0.75, 0.0}

\begin{document}

 \title{Scattering between minivalleys in a moir\'{e} material}

\author{Petar Tomić}
\email{ptomic@phys.ethz.ch}
\author{Peter Rickhaus}
\affiliation{Laboratory for Solid State Physics, ETH Z\"{u}rich, CH-8093 Z\"{u}rich, Switzerland}
\author{Aitor Garcia-Ruiz}
\affiliation{National Graphene Institute, University of Manchester, Manchester M13 9PL, United Kingdom}
\author{Giulia Zheng}
\author{El\'{i}as Portol\'{e}s}
\affiliation{Laboratory for Solid State Physics, ETH Z\"{u}rich, CH-8093 Z\"{u}rich, Switzerland}
\author{Vladimir Fal'ko}
\affiliation{National Graphene Institute, University of Manchester, Manchester M13 9PL, United Kingdom}
\affiliation{Henry Royce Institute for Advanced Materials, M13 9PL, Manchester, UK}
\author{Kenji Watanabe}
\author{Takashi Taniguchi}
\affiliation{National Institute for Material Science, 1-1 Namiki, Tsukuba 305-0044, Japan}
\author{Klaus Ensslin}
\author{Thomas Ihn}
\author{Folkert K. de Vries}
\email{devriesf@phys.ethz.ch}
\affiliation{Laboratory for Solid State Physics, ETH Z\"{u}rich, CH-8093 Z\"{u}rich, Switzerland}


\begin{abstract}
A unique feature of the complex band structures of moiré materials is the presence of minivalleys, their
hybridization, and scattering between them.
Here we investigate magneto-transport oscillations caused by scattering between minivalleys - a phenomenon analogous to magneto-intersubband oscillations - in a twisted double bilayer graphene sample with a twist angle of $\SI{1.94}{^\circ}$.
We study and discuss the potential scattering mechanisms and find an electron-phonon mechanism and valley conserving scattering to be likely.
Finally, we discuss the relevance of our findings for different materials and twist angles.
\end{abstract}

\maketitle

Two-dimensional moir\'{e} materials are formed by stacking van der Waals materials such that the layers couple and an in-plane superlattice emerges. 
The superlattice formed depends on the twist (and lattice mismatch) between the layers.
Graphene is a typical van der Waals material and has a honeycomb lattice build up from two hexagonal sublattices. The wavefunctions of the two valleys $K$ and $K'$ in reciprocal space are sublattice polarized at the Dirac point~\cite{CastroNeto_2009}. Similarly, for trigonal moir\'{e} lattices, such as twisted graphene~\cite{Kuwabara_1990,Bistritzer_2011}, the wavefunctions of the minivalleys $\kappa$ and $\kappa'$ are polarized on the twisted layers when the interlayer coupling is weak. Even though this is a generic property of trigonal moir\'{e} lattices, little is known about the scattering of charge carriers between these minivalleys. 


We focus on twisted graphene since recently a plethora of correlated states has been observed~\cite{Cao_2018_insulator,Cao_2018_superconductor,Sharpe2019,Lu2019,Burg_2019,Cao_2020_TDBG,Rickhaus_2020_CDW}, and in view of these correlations, scattering in twisted graphene is a highly interesting topic.
Specifically, we choose to work with twisted double bilayer graphene (TDBG) with weak coupling between the layers, because this offers excellent control over the minivalley occupation, and high quality electron transport.
This system resembles that of a weakly coupled double quantum well, where the two minivalleys around $\kappa$ and $\kappa'$ play the roles of the two subbands. Since the wavefunctions of the minivalleys are mostly bilayer polarized, a dual gate geometry provides independent control over the density in the two minivalleys, and with that, their energetic alignment~\cite{deVries_2020}.

 

A common way to obtain the leading scattering mechanism is analyzing the temperature dependence of the electrical resistivity~\cite{Bolotin_2008}.
Measurements of the resistivity in magic-angle twisted bilayer graphene have shown a linear temperature dependence, suggestive of electron-phonon scattering~\cite{Polshyn_2019,Wu_2019} or strange metallic behavior~\cite{Cao_2020,Hwang_2019}.
As the scattering between minivalleys is not necessarily the leading scattering mechanism, we introduce a more targeted approach.
When applying a magnetic field, Landau levels are formed in both minivalleys separately. 
The energetic (mis)alignment of the modulated densities of states in the two minivalleys leads to oscillations in the interminivalley scattering. 
This oscillating interminivalley scattering is reflected in electrical transport through an effect analogous to magneto intersubband oscillations (MISO)~\cite{polyanovskii1988anomalous,dmitriev2012nonequilibrium}. In the following, we will refer to these magneto inter-minivalley oscillations as MISO, since they encapsulate the same physical phenomenon.
The method we present is transferable to other moir\'{e} materials with a well developed Landau level spectrum as well as control of the energetic alignment of the minivalleys.


Here, we report measurements of MISO in TDBG with a twist angle of $1.94^\circ$ and investigate the interminivalley scattering in the regime of hole-like states ($n<0$). 
We introduce the bandstructure and tunability of the TDBG device by analyzing its Shubnikov-de Haas oscillations (SdHO) as a function of density and displacement field. 
Then we investigate two regions where MISO are particularly pronounced.
First, towards the Lifshitz transition~\cite{Lifshitz_1960}, we study MISO as a function of temperature and displacement field and discuss the scattering mechanism. 
Second, at the onset of the second minivalley, we discuss implications of the underlying scattering mechanism based on the observed valley degeneracy lifting.
Finally, we discuss the impact of our findings with regard to different materials and twist angles.

\begin{figure*}[t]
    \centering
    \includegraphics{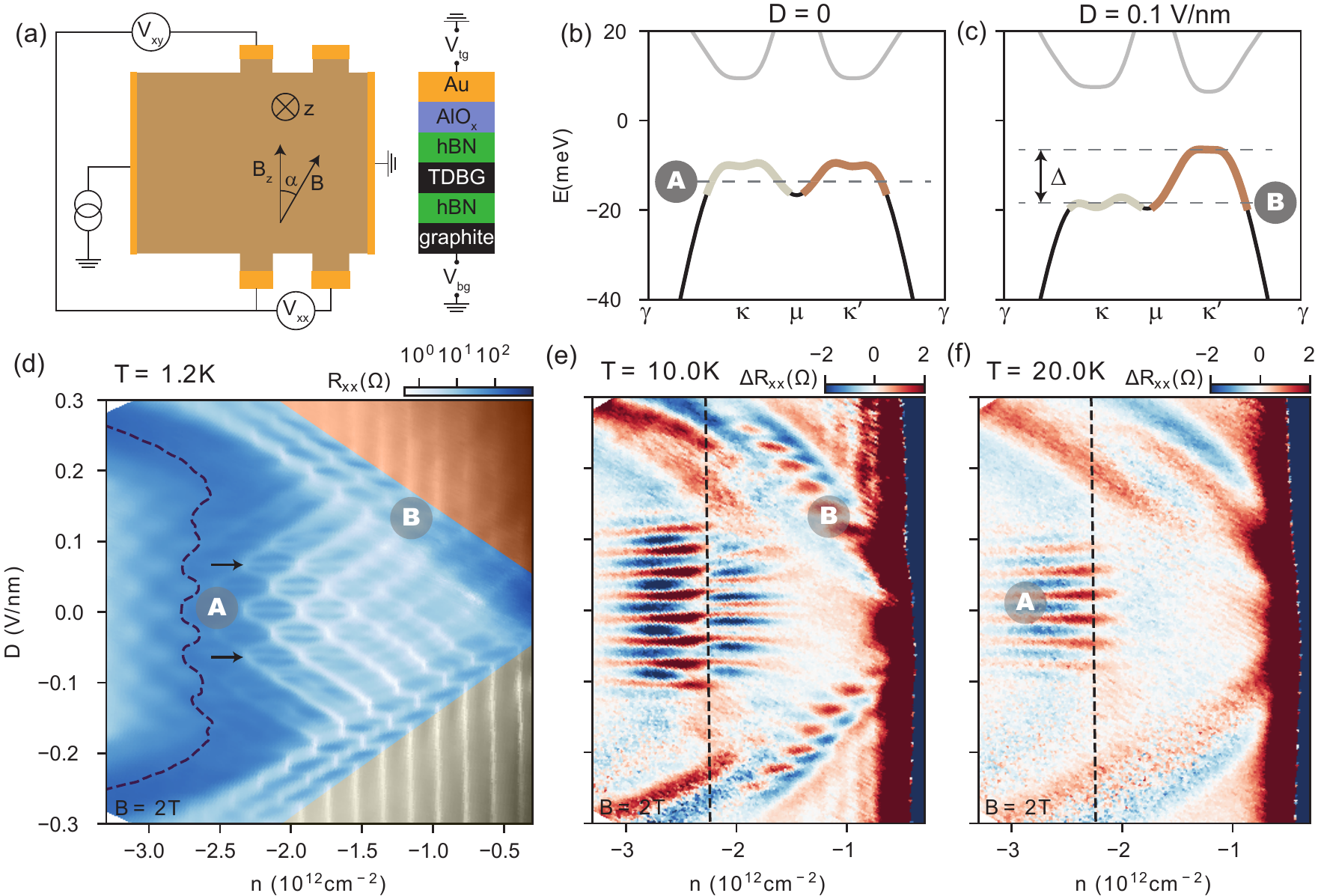}
    \caption{(a) Schematic overview of the device, including setup to measure the longitudinal and Hall voltage ($V_\mathrm{xx}$ and $V_\mathrm{xy}$), and the alignment of magnetic field $B$. Cross-section of the device featuring the graphite, hexagonal boron nitride (hBN), TDBG, aluminium-oxide (AlO$_\mathrm{x}$), and gold (Au) layers, and the top and bottom gates ($\Vtg$ and $\Vbg$) indicated.
    (b,c) Calculated band structure of TDBG ($\SI{1.94}{^\circ}$) for displacement field $D=0$ and $D=\SI{0.1}{V/nm}$, respectively. Minivalleys $\kappa$ and $\kappa'$ are indicated as well as the regions A and B.
    (d) $\Rxx$ as a function of density $n$ and $D$ at temperature $T=\SI{1.2}{K}$ and $B=\SI{2}{T}$. Regions A and B are indicated. The overlayed colors bronze and gray indicate single minivalley regimes. The LT is highlighted by the dashed line.
    (e,f) Resistance modulation $\Delta\Rxx(n,D)$ at $B=\SI{2}{T}$ and $T=\SI{10}{K}$ and $T=\SI{20}{K}$, respectively. The density used in Figs.~\ref{fig:fig2} and \ref{fig:fig3} is indicated with a black dashed line.}
    \label{fig:fig1}
\end{figure*}

We fabricate a Hall bar device [Fig.~\ref{fig:fig1}(a)] from TDBG sandwiched between two hexagonal boron nitride layers.
A four-terminal current-bias setup is used to obtain the longitudinal resistance $\Rxx$ and Hall resistance $\Rxy$, utilizing standard lock-in techniques with $I_\mathrm{AC}=\SI{100}{nA}$.
Crucially, we are able to control the density $n$ and displacement field $D$ separately in the device through voltages applied to the top gate ($\Vtg$) and bottom gate ($\Vbg$) (see Supplementary Material~\cite{SM}).
We investigate the device in a $\mathrm{^4He}$ cryostat with a base temperature of $T=\SI{1.2}{K}$, and implemented temperature control.
The device is identical to the one investigated in Ref.~\cite{twistersBZ_2021}. For details on the device fabrication and data analysis see the Supplementary Material~\cite{SM}.



The bandstructure of TDBG with a twist angle of $1.94^\circ$ is presented in Fig.~\ref{fig:fig1}(b). The bands in the mini-Brioullin zone show local minima, maxima and band gaps around the two minivalleys $\kappa$ and $\kappa'$. Without an applied displacement field D the minivalleys are occupied equally (region A). In contrast, when a displacement field is applied, the energetic alignment of the minivalleys $\Delta$ is altered~\cite{deVries_2020}. Independently tuning the Fermi energy now allows one to only occupy the bands centered around $\kappa'$ [Fig.~\ref{fig:fig1}(c)], and reach asymmetric minivalley occupations (region B).


The tunability of our device with respect to the occupation of different minivalleys in the hole bands is investigated using SdHO. 
In Fig.~\ref{fig:fig1}(d) we show $\Rxx$ as a function of displacement field $D$ and total density $n$ measured at constant magnetic field $B=\SI{2}{T}$ and temperature $T=\SI{1.2}{K}$. 
We observe a single set of SdHO in the regions masked with bronze and gray in  Fig.~\ref{fig:fig1}(d), indicating the occupation of a single minivalley.
This corresponds to the Fermi energy being tuned into the band gap of either minivalley [Fig.~\ref{fig:fig1}(c)].
The blue region in Fig.~\ref{fig:fig1}(d) corresponds to the configuration where both minivalleys $\kappa$ and $\kappa'$ are occupied, giving rise to a pattern of two sets of SdHO. 
We reproduce this pattern within a simple model by considering screening effects of the bilayers as presented in the Supplementary Material~\cite{SM}.
Finally, we highlight the Lifshitz transition~\cite{Lifshitz_1960} in Fig.~\ref{fig:fig1}(d) by drawing the contour (dashed line) where $\Rxy=0$. At this line, the topology of the Fermi surface changes as the Fermi energy crosses the saddle point in the bandstructure at the $\mu$ point [see Fig.~\ref{fig:fig1}(b,c)] ~\cite{deVries_2020}.

Interestingly, apart from the SdHO pattern, density independent resistance minima cutting through the middle of the hexagons are observed indicated with black arrows in the region labeled A in Fig.~\ref{fig:fig1}(d).
Unlike SdHO, which are strongly thermally damped, these resistance minima are more pronounced at higher temperatures as seen at $T=\SI{10}{K}$ and $T=\SI{20}{K}$ in Fig.~\ref{fig:fig1}(e,f), where a smooth background has been subtracted for improved clarity of the presentation.
These oscillations are MISO~\cite{polyanovskii1988anomalous,dmitriev2012nonequilibrium}, caused by an oscillating interminivalley scattering rate.
The displacement field tunes the energy offset between the minivalleys, $\Delta$ (indicated in Fig~\ref{fig:fig1}(c)) that periodically changes the energetic alignment of the Landau levels in the minivalleys. This results in a periodic modulation of the interminivalley scattering rate that we observe as MISO in Fig.~\ref{fig:fig1}(d-f).

The unprecedented tunability of our TDBG device allows us to study MISO at various possible relative energetic alignments of the minivalleys. 
In particular, we observe enhanced MISO in two distinct regions denoted A and B in Fig.~\ref{fig:fig1}(d-f).
Region A is located towards the Lifshitz transition for moderate values of displacement fields, and region B at the onset of the second minivalley.

\begin{figure}[t]
    \centering
    \includegraphics[width=0.5\textwidth]{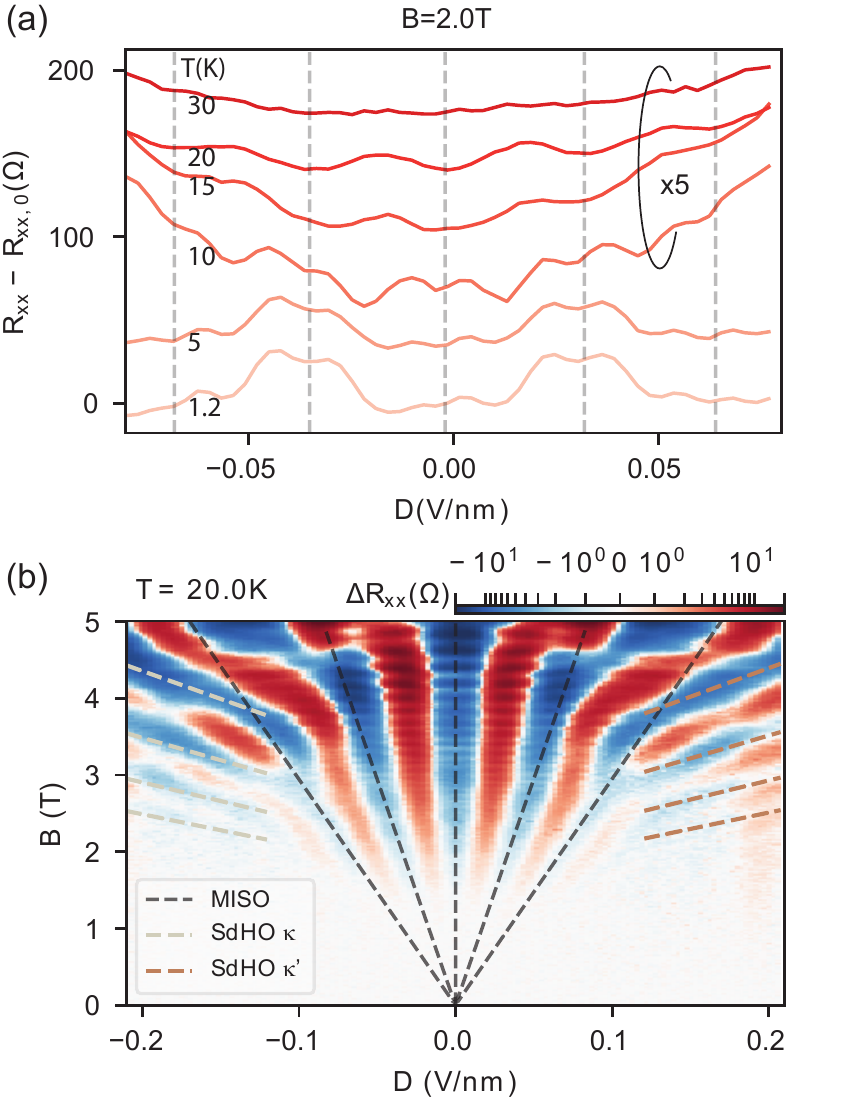}
    \caption{(a) Line traces of the $\Rxx-R_\mathrm{xx,0}$ (value of $\Rxx$ at $D=0$) versus displacement field $D$ for various temperatures $T$ as indicated, at constant $n=\SI{-2.25e12}{cm^{-2}}$ and $B=\SI{2}{T}$. The traces are offset by \SI{40}{\Omega} and the upper four traces are multiplied by 5 for clarity. Vertical dashed lines mark positions of MISO as guides to the eye. (b) Resistance modulation $\Delta \Rxx$ as a function of $D$ and $B$, taken at $n=\SI{-2.25e12}{cm^{-2}}$ and $T=\SI{20}{K}$. The black dashed lines represent the result of the basic MISO model, while gray and bronze dashed lines represent SdHO originating from $\kappa$ and $\kappa'$ minivalley respectively (see Supplementary Material~\cite{SM}).}
    \label{fig:fig2}
\end{figure}

We start discussing region A showing unambiguously the difference between MISO and SdHO.
Line traces of $\Rxx$ as a function of $D$ measured at different temperatures and constant density $n=\SI{-2.25e12}{cm^{-2}}$ are shown in Fig.~\ref{fig:fig2}(a) revealing the suppression of SdHO with temperature, while MISO persist.
At $T=\SI{20}{K}$, as in Fig.~\ref{fig:fig1}(f), only MISO are left, where as at $T=\SI{1.2}{K}$, as in Fig.~\ref{fig:fig1}(d), SdHO are dominant and MISO are slightly visible as well.

In order to show that the oscillation spectrum of MISO also differs from SdHO we present the resistivity modulation $\Delta\Rxx$ as a function of $B$ and $D$ at $T=\SI{20}{K}$ in Fig.~\ref{fig:fig2}(b).
Note, displacement field independent Azbel-Brown-Zak oscillations (ABZO)~\cite{KrishnaKumar_2017,twistersBZ_2021} are present in the whole magnetic field range.
For comparison, we plot the results of a basic MISO model (see Supplementary Material~\cite{SM}) using dashed lines that highlight the alignment of Landau levels from different minivalleys.
The condition is fulfilled when the energy spacing of an integer number of Landau levels fits the energy offset between the minivalleys, i.e. when $\Delta (D) = k \hbar e B/m_\mathrm{eff}$ where $k$ is an integer and $m_\mathrm{eff} = 0.06 m_\mathrm{e}$ the effective mass.
At $T=\SI{20}{K}$ we estimate that only a few Landau levels participate in transport.
The model fits well for moderate $D$ and deviates at higher $D$, where we suspect the alignment condition is broken because $m_\mathrm{eff}$ in the minivalleys becomes dissimilar. This is in contrast to semiconductors with mostly parabolic bandstructure, where larger numbers of Landau levels can overlap at the same time. 
Finally, we like to point out the MISO fan in Fig.~\ref{fig:fig2}(b) is different from SdHO Landau fans since it has its origin at $D=0$ [not at finite $D$ as for SdHO in Supplemental Fig. 7], and it fans out in both directions in $D$.

Generally MISO is considered to be caused by impurity scattering~\cite{dmitriev2012nonequilibrium}. One would therefore expect to see maxima in the resistivity when the Landau levels from both minivalleys are aligned. Here, we observed the opposite. 
Observing minima at MISO resonances indicates that phonon-assisted instead of impurity intersubband scattering is the dominant mechanism~\cite{raichev2010magnetoresistance}. 
We expect the electron-phonon scattering to be quasi-elastic, because in Fig.~\ref{fig:fig2}(b) the experimentally observed phase of the oscillation matches with the phase of the Landau level alignment in the basic MISO model.
Since we observe the MISO resonances as resistance minima in the full temperature range [Fig.~\ref{fig:fig2}(a)], we envision scattering by low energy phonons (e.g. flexural phonons) that have a flat dispersion~\cite{Koshino_2019, Cocemasov_2013}.

Additionally, we measured the temperature dependence of the resistivity at $B=0$ and $D=0$, which shows a linear behavior, indicative of an electron-phonon mechanism. 
However, since the amplitude of the MISO is small compared to the smooth background [e.g. $\Delta \Rxx/R_\mathrm{xx,0}\sim 1\%$ at $T=\SI{20}{K}$ in Fig.~\ref{fig:fig2}(a)], the interminivalley scattering is probably not the dominant scattering mechanism determining the resistance. In the Supplementary Material~\cite{SM}, apart from a detailed analysis of the temperature-dependence of the data, we present a theoretical model that reveals that intraminivalley electron-phonon scattering is likely to be the dominant mechanism.

\begin{figure}[t]
    \centering
    \includegraphics{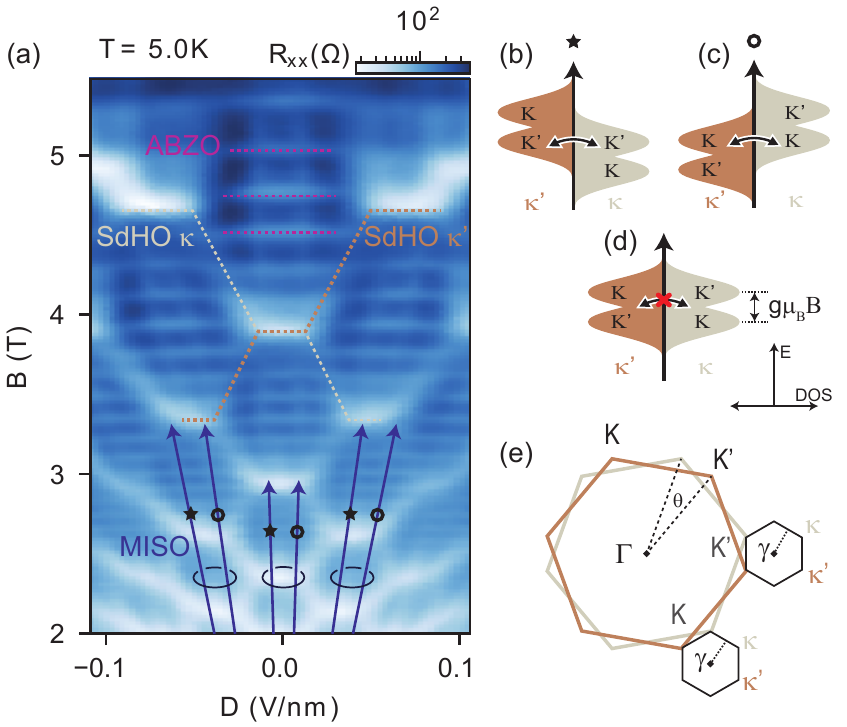}
    \caption{(a) $\Rxx(D,B)$ at $T=\SI{5}{K}$ and $n=\SI{-2.25e12}{cm^{-2}}$. We highlight an example of SdHO from $\kappa$ and $\kappa'$ carriers with gray and bronze dashed lines, several ABZO by purple dashed lines, and the split MISO peaks with the dark blue arrows.
    (b,c,d) Possible description of splitting of the MISO peaks. The density of states (DOS) for both minivalleys $\kappa$ and $\kappa'$ are shown for $D<0$, $D>0$ and $D=0$, respectively. The valley splitting and proposed scattering is indicated.
    (e) Mini-Brillouin zones of TDBG with labeled valleys ($K$ and $K'$) and minivalleys ($\kappa$ and $\kappa'$).}
    \label{fig:fig3}
\end{figure}

We continue by analyzing MISO as a function of displacement and magnetic field at $T=\SI{5}{K}$ in Fig.~\ref{fig:fig3}(a), focusing on the same density as in Fig.~\ref{fig:fig2}(b).
The hexagonal pattern is formed by two sets of SdHO originating from respective minivalleys, as indicated in Fig.~\ref{fig:fig3}(a), whose discontinuous structure is a result of screening effects.
Additionally, displacement field independent ABZO are present throughout the whole magnetic field range.
Interestingly, instead of single resistance minima, as observed at the alignment condition of MISO in Fig.~\ref{fig:fig2}(b), the MISO resonances are split into two resistance minima at higher magnetic fields, as highlighted by the dark blue arrows in Fig.~\ref{fig:fig3}(a), while the SdHO are not split.
We extract an effective Land\'{e} $g$-factor of $g\approx5$, pointing towards a splitting of the valley ($K$ and $K'$) degeneracy rather than the spin~\cite{Lee_2020}.

The possible Landau level alignments taking into account valley splitting of a single Landau level in each minivalley are schematically shown in Fig.~\ref{fig:fig3}(b-d), where three typical energetic offsets are sketched.
From this picture one would expect to observe three MISO minima. However, since we only observe two (configuration b and c), the interminivalley scattering must have a valley selection criterion.
Note that the MISO splitting is twice that of the Landau levels, which makes it plausible that the splitting is not apparent in the SdHO.
In Fig.~\ref{fig:fig3}(e) we sketch the mini-Brillouin zones with respective valleys and minivalleys. Considering the relative distance in $k$-space it is more likely that the scattering is valley conserving.
In combination with a quasi-elastic scattering mechanism, this implies that the valley splitting in the two graphene bilayers should be opposite and scattering is only allowed when the same valleys line up, as sketched in Fig.~\ref{fig:fig3}(b,c). 
Future theoretical work is needed to confirm this possible mechanism.

\begin{figure}[t]
    \centering
    \includegraphics{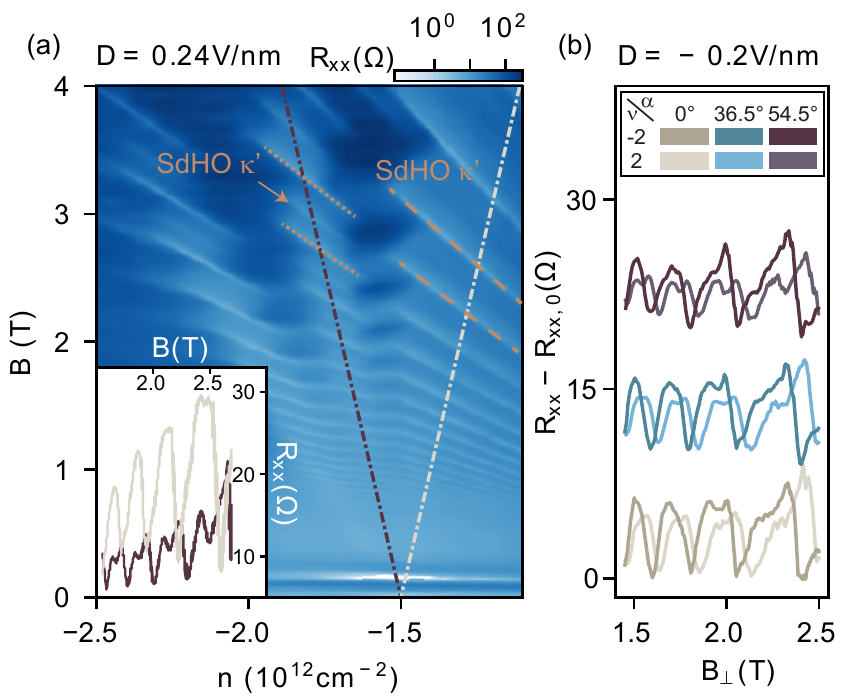}
    \caption{(a) $\Rxx(n,B)$ taken at $D=\SI{0.24}{V/nm}$ and $T=\SI{1.2}{K}$. The inset shows $\Rxx$ as a function of $B$, for constant second minivalley filling factors $\nu=\pm2$, as indicated by the dotted lines. 
    (b) $\Rxx-R_\mathrm{xx,0}$ taken at $D=\SI{-0.2}{V/nm}$ as a function of perpendicular component of the magnetic field $B_{\bot}$ for $\nu=\pm2$ and  $\alpha = 0^\circ, 36.5^\circ, 54.5^\circ$}
\label{fig:fig4}
\end{figure}


After analyzing the region of enhanced MISO in region A of Fig.~\ref{fig:fig1}(d), we shift our attention to the region at the onset of the second minivalley occupation, denoted by B in Fig.~\ref{fig:fig1}(d).
In Fig.~\ref{fig:fig4}(a) we plot $\Rxx$ as a function of $B$ and $n$, at finite displacement field and temperature of $T=\SI{1.2}{K}$. 
At densities below $n\sim \num{-1.5e12}cm^{-2}$, which is the density onset of the $\kappa$ minivalley, we measure the Landau fan of $\kappa'$ minivalley highlighted with bronze dashed lines in Fig.~\ref{fig:fig4}(a).
Once the Landau levels of the second minivalley appear, the levels in the already occupied $\kappa'$ minivalley (dotted bronze lines) split as indicated with the bronze arrow in Fig.~\ref{fig:fig4}(a), signalling a transition from 4-fold to 2-fold degenerate levels [see also inset of Fig.~\ref{fig:fig4}(a)]. 
To find out which degeneracy (valley or spin) is lifted, we perform additional measurements, using a different cryostat, for two different angles of the magnetic field with respect to the normal of device [see Fig.~\ref{fig:fig4}(b)].
Since the splitting depends on $B_{\bot}$ (and not on $|B|$), we speculate that the valley ($K$ and $K'$) degeneracy in the already occupied $\kappa'$ minivalley is lifted.

The lifting of the valley degeneracy once the second minivalley gets occupied can be a result of electron-electron interactions, as it is not expected to occur due to single particle effects~\cite{Crosse_2020}. A possible electron-electron interaction at play is the valley exchange effect. Furthermore, due to high effective mass at the onset of the second minivalley, screening effects can improve the mobility leading to better line-width of SdHO hence revealing the splitting in the spectrum. However, since we observe enhanced MISO, i.e. stronger scattering, at the onset of the second minivalley, we speculate that the observed enhanced inter-minivalley scattering leading to MISO in region B of Fig.~\ref{fig:fig1}(d) is due to an electron-electron scattering mechanism.


In summary, by using MISO as a targeted measurement approach, we have investigated interminivalley scattering in a moir\'{e} material. We found two regions of enhanced MISO, and discussed the likely scattering mechanisms of electron-phonon scattering with a valley selection rule in the vicinity of the Lifshitz transition and electron-electron scattering at the onset of the second minivalley.
The described measurement technique is transferable to other moir\'{e} materials~\cite{Masseroni_2021}, given that they adhere to the prerequisites of a clear Landau level spectrum and sufficient tunability.
These observations may give a handle to refine theoretical models that aim to capture interactions in moir\'{e} systems, such as magic-angle twisted bilayer graphene.

The data used in this Letter will be made available through the ETH Research Collection.

\begin{acknowledgements}
We acknowledge the support from Peter Maerki, Thomas Baehler and the staff of the ETH FIRST cleanroom facility.
We acknowledge support from the Graphene Flagship, the Swiss National Science Foundation via NCCR Quantum Science and from the European Union's Horizon 2020 research and innovation programme under grant agreement No 862660/QUANTUM E LEAPS.
E.P acknowledges support of a fellowship from ”la Caixa” Foundation (ID 100010434) under fellowship code LCF/BQ/EU19/11710062.
K.W. and T.T. acknowledge support from the Elemental Strategy Initiative conducted by the MEXT, Japan, Grant Number JPMXP0112101001,  JSPS KAKENHI Grant Number JP20H00354 and the CREST(JPMJCR15F3), JST.
\end{acknowledgements}


%

\end{document}


\title{Supplementary Material for Scattering between minivalleys in a moir\'{e} material}

\author{Petar Tomić}
\email{ptomic@phys.ethz.ch}
\author{Peter Rickhaus}
\affiliation{Laboratory for Solid State Physics, ETH Z\"{u}rich, CH-8093 Z\"{u}rich, Switzerland}
\author{Aitor Garcia-Ruiz}
\affiliation{National Graphene Institute, University of Manchester, Manchester M13 9PL, United Kingdom}
\author{Giulia Zheng}
\author{El\'{i}as Portol\'{e}s}
\affiliation{Laboratory for Solid State Physics, ETH Z\"{u}rich, CH-8093 Z\"{u}rich, Switzerland}
\author{Vladimir Fal'ko}
\affiliation{National Graphene Institute, University of Manchester, Manchester M13 9PL, United Kingdom}
\affiliation{Henry Royce Institute for Advanced Materials, M13 9PL, Manchester, UK}
\author{Kenji Watanabe}
\author{Takashi Taniguchi}
\affiliation{National Institute for Material Science, 1-1 Namiki, Tsukuba 305-0044, Japan}
\author{Klaus Ensslin}
\author{Thomas Ihn}
\author{Folkert K. de Vries}
\email{devriesf@phys.ethz.ch}
\affiliation{Laboratory for Solid State Physics, ETH Z\"{u}rich, CH-8093 Z\"{u}rich, Switzerland}

\maketitle

\onecolumngrid
\section{Band structure of twisted double bilayer graphene under external displacement fields}
Twisted double bilayer graphene (tDBLG) consists of two Bernal-aligned graphene bilayers stacked on top of each other with their rotational axes tilted by and angle $\theta$. Here we consider both layers to be AB stacked, which is often referred to as ABAB-tDBLG. To construct the Hamiltonian of the system, both aligned and twisted interfaces are treated using the Slonczewski-Weiss-McClure (SWMcC) Hamiltonian \cite{Slonczewski1958,McClure1957,McClure1960}, and the continuum model \cite{Lopes2007,Bistritzer_2011}, to describe the aligned and twisted interfaces, respectively. In the basis of sublattice Bloch states, $\left\{\Psi_{A1},\Psi_{B1},\dots,\Psi_{A4},\Psi_{B4}\right\}$ the total Hamiltonian of twisted double bilayer graphene under a displacement field takes the general form
\begin{align}\label{eq:Total_Hamiltonian}
\hat{H}=&
\left(
\begin{matrix}
\hat{H}_{BA}^{\mathrm{t}}&\hat{\mathcal{T}}\\
\hat{\mathcal{T}}^{\dagger}&\hat{H}_{BA}^{\mathrm{b}}
\end{matrix}
\right)+\hat{\mathcal{E}}.
\end{align}
In the first matrix, the diagonal $4\times4$ blocks describe the coupling within the aligned bilayers, and can be written as
\begin{align}
\hat{H}_{BA}^\lambda=&
\left(
\begin{matrix}
0&v\pi_{\xi,\lambda}^*&-v_4\pi_{\xi,\lambda}^*&-v_3\pi_{\xi,\lambda}\\
v\pi_{\xi,\lambda}&\Delta'&\gamma_1&-v_4\pi_{\xi,\lambda}^*\\
-v_4\pi_{\xi,\lambda}&\gamma_1&\Delta'&v\pi_{\xi,\lambda}^*\\
-v_3\pi_{\xi,\lambda}^*&-v_4\pi_{\xi,\lambda}&v\pi_{\xi,\lambda}&0
\end{matrix}
\right),\qquad \lambda=\mathrm{t,b}
\end{align}
with $\pi_{\xi,t/b}\equiv \xi p_x+i(p_y\mp \theta K)$, $(p_x,p_y)$ is the momentum measured from the corner of the Brillouin zone, and $(v,v_3,v_4)=(1.02,0.12,0.05)10^{6}\,\mathrm{m/s}$ and $(\gamma_1,\Delta')=(381,22)\,\mathrm{meV}$ \cite{Kuzmenko2009} are the SWMcC parameters. In contrast, the off-diagonal $4\times4$ blocks capture the interlayer hopping processes across the twisted interface, and takes the form
\begin{align}
\hat{\mathcal{T}}=&
\sum_{j=0}^2
\left(
\begin{matrix}
\hat{0}_2&\hat{0}_2\\\hat{T}_j&\hat{0}_2
\end{matrix}
\right)
\delta_{\boldsymbol{k}',\boldsymbol{k}+\boldsymbol{G}_j},\qquad
\hat{T}_j=
\frac{\gamma_1}{3}
\left(
\begin{matrix}
1&e^{i\xi\frac{2\pi}{3}j}\\
e^{-i\xi\frac{2\pi}{3}j}&1
\end{matrix}
\right),\nonumber
\end{align} 
where $\hat{0}_2$ is the $2\times2$ empty matrix, and $\boldsymbol{G}_j=
\xi \theta K \left[-\sin({2\pi j}/{3}),1-\cos({2\pi j}/{3})\right]$. The second matrix in Eq. (\ref{eq:Total_Hamiltonian}), $\hat{\mathcal{E}}$, captures the effect of the electric fields between the layers, induced by both external displacement fields and charge redistribution. This matrix can be expressed as a function the energy difference between two consecutive layers, $\Delta_{j,j+1}$, as follows
\begin{align}
\hat{\mathcal{E}}=
\left(
\begin{matrix}
\left(
\Delta_{3,4}+\Delta_{2,3}+\Delta_{1,2}
\right)\hat{\mathbb{1}}_2&\hat{0}_2&\hat{0}_2&\hat{0}_2\\
\hat{0}_2&
\left(\Delta_{2,3}+\Delta_{1,2}
\right)\hat{\mathbb{1}}_2&\hat{0}_2&\hat{0}_2\\
\hat{0}_2&\hat{0}_2&\Delta_{1,2}\hat{\mathbb{1}}_2&\hat{0}_2\\
\hat{0}_2&\hat{0}_2&\hat{0}_2&\hat{0}_2
\end{matrix}
\right),
\end{align}
where $\hat{\mathbb{1}}_2$ is the $2\times2$ unit matrix and we set the on-site energy of the bottommost layer to zero. For a given perpendicular displacement field, $D_z$, the three parameters $\Delta_{j,j+1}$ are obtained using \cite{Slizovskiy2019}
\begin{subequations}\label{eq:Deltas}
\begin{align}
\Delta_{1,2}=&
\frac{eD_zc_0}{\epsilon_0\epsilon_z}+
\frac{e^2c_0(n_2-n_1)}{2\epsilon_0}
\frac{1+\epsilon_z^{-1}}{2}+
\frac{e^2c_0(n_3+n_4)}{2\epsilon_0\epsilon_z},\\
\Delta_{2,3}=&
\frac{eD_zc_0}{\epsilon_0\epsilon_z}+
\frac{e^2c_0(n_3-n_2)}{2\epsilon_0}
\frac{1+\epsilon_z^{-1}}{2}+
\frac{e^2c_0(n_4-n_1)}{2\epsilon_0\epsilon_z},\\
\Delta_{3,4}=&
\frac{eD_zc_0}{\epsilon_0\epsilon_z}+
\frac{e^2c_0(n_4-n_3)}{2\epsilon_0}
\frac{1+\epsilon_z^{-1}}{2}-
\frac{e^2c_0(n_1+n_2)}{2\epsilon_0\epsilon_z}.
\end{align}
\end{subequations}

Above, $c_0\approx3.35\,\mathrm{\AA}$ denotes the interlayer distance, $\epsilon_z\approx2.65$ is the dielectric susceptibility of graphene in the perpendicular direction \cite{Slizovskiy2019} and the layer densities $n_i$ are obtained integrating over the first mini Brillouin zone (mBZ),

\begin{align}\label{eq:densities}
n_{j}=4
\int_{\mathrm{mBZ}}
\frac{\mathrm{d}\boldsymbol{p}}{(2\pi\hbar)^2}
\sum_{\beta}
\left[
\left(
|\Psi_{Aj,\boldsymbol{p}}^{\beta}|^2+
|\Psi_{Bj,\boldsymbol{p}}^{\beta}|^2
\right)
\Theta(E_F-E_{\boldsymbol{p}}^{\beta})
-\frac{1}{8}
\right],
\end{align}
where the 4 accounts for the spin and valley degeneracy, $\Psi_{\lambda j,\boldsymbol{p}}^{\beta}$ is the amplitude in sublattice $\lambda=A,B$ of the layer $j$ of the eigenvector associated to the band $\beta$ and momentum $\boldsymbol{p}$, the Heaviside function ensures the integration up to the Fermi level $E_F$ and the factor $1/8$ is needed to account for the compensating charge of carbon ions. The evaluation of the interlayer energy differences in Eqs. (\ref{eq:Deltas}) is carried out self-consistently. We use $\Delta_{j,j+1}^{(0)}=eD_zc_0/\epsilon_0\epsilon_z$ as initial guess values for all interlayer energy differences, calculate the resulting layer densities in Eq. (\ref{eq:densities}) and insert them into Eq.(\ref{eq:Deltas}) to obtain the following values. The process is iterated convergence is reached. 

\begin{figure}
\begin{center}
\includegraphics[width=0.8\columnwidth]{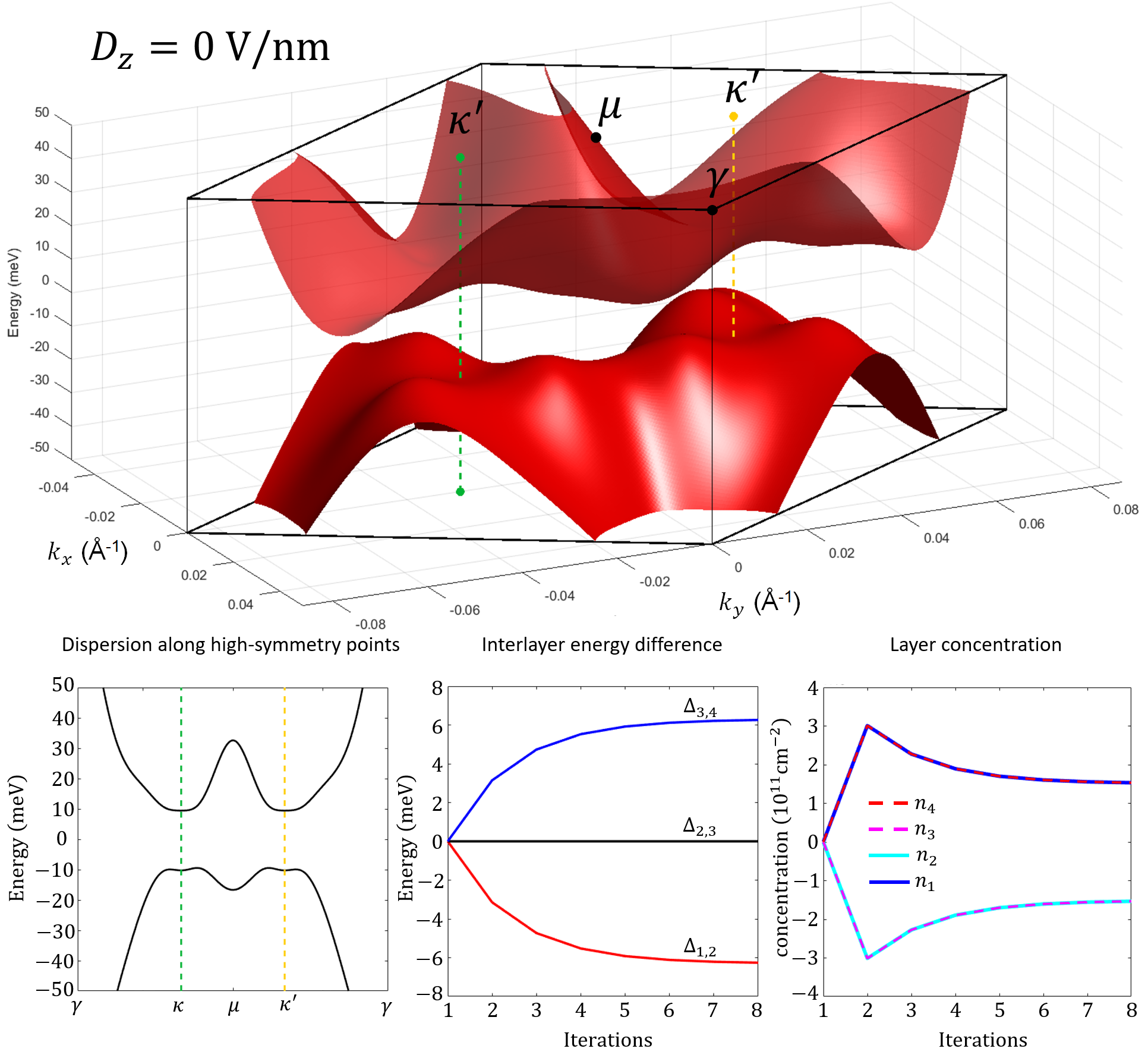}
\caption{ (Top) First conduction and valence bands of twisted double bilayer graphene (ABAB) in the first mini Brillouin zone, without any external displacement and with total density $n=\sum_jn_j=0$. (Bottom-left) Dispersion along the high symmetry points, (Bottom-middle) convergence check for the interlayer energy difference and (Bottom-right) convergence check for the layer concentration. \label{fig:dispersion_without}}
\end{center}
\end{figure}

\begin{figure}
\begin{center}
\includegraphics[width=0.8\columnwidth]{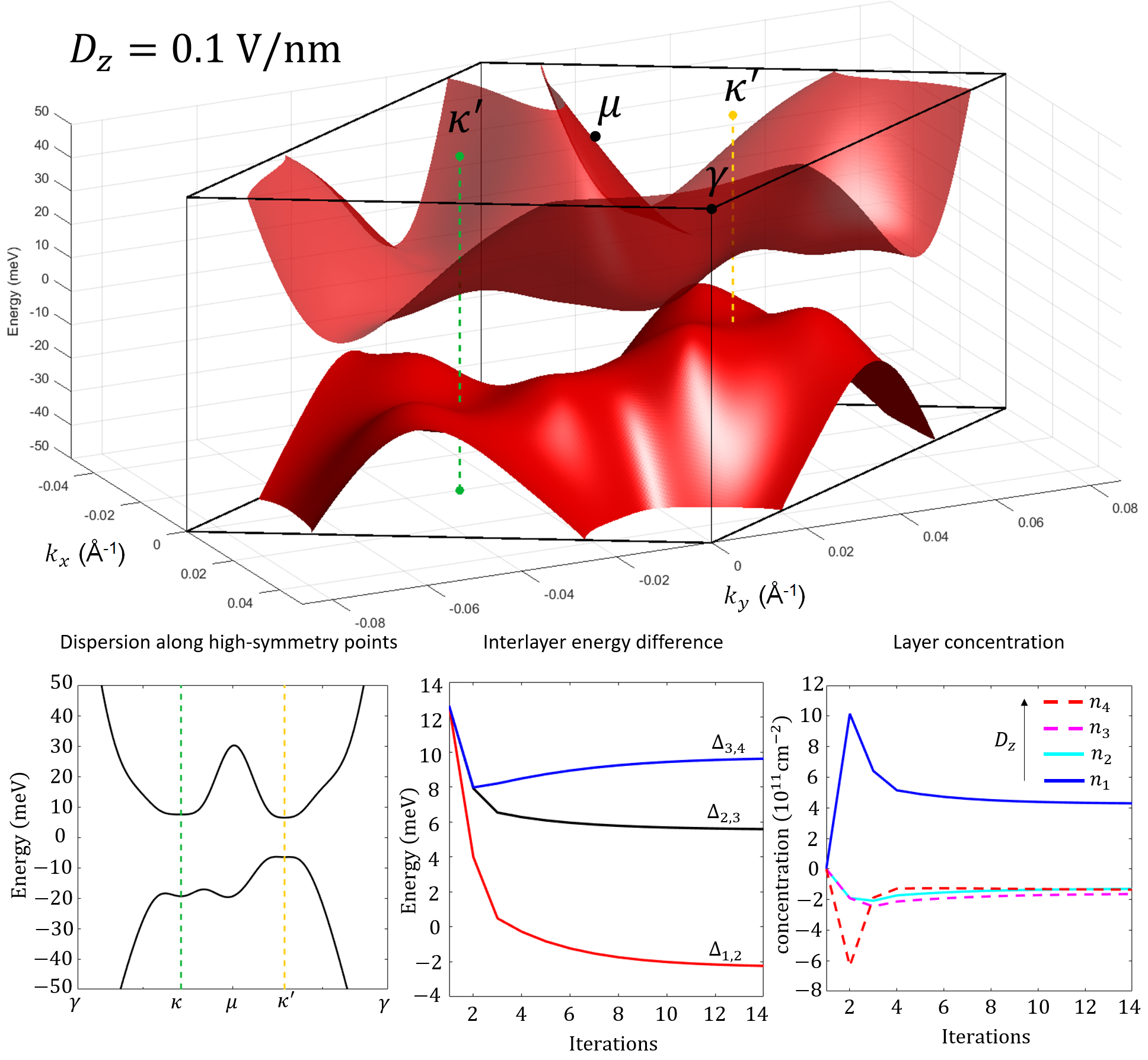}
\caption{ (Top) First conduction and valence bands of twisted double bilayer graphene (ABAB) in the first mini Brillouin zone under the application of a displacement field of $D_z=0.1$ V/nm and total density $n=\sum_jn_j=0$. (Bottom-left) Dispersion along the high symmetry points, (Bottom-middle) convergence check for the interlayer energy difference and (Bottom-right) convergence check for the layer concentration. \label{fig:dispersion_with}}
\end{center}
\end{figure}

We apply the method described above to describe the band structure of ABAB-tDBLG with a twist angle of $\theta=1.94^\circ$ for $D_z=0$ V/nm, and results are shown in Fig. \ref{fig:dispersion_without}. We observe that charge redistribution occurs, even without the presence of an external displacement field, due to the different interlayer potentials across the three interfaces. In particular, charge redistributes to the outermost layers, which leaves an equal concentration, but with opposite sign, in the inner layers. This charge redistribution generates a small interlayer energy difference across the Bernal interfaces of about $\sim+6$ meV and $\sim-6$ meV in the top and bottom bilayers, respectively. Upon the application of an external displacement field of $D_z=0.1$ V/nm, the electronic charge redistributes from the top three layers to the bottommost layer (see Fig. \ref{fig:dispersion_with}). The gaps across the Bernal interfaces increases by about $~4$ meV, while the gap in the twisted interface increases by $~6$ meV. As a result, the gap at the $\kappa'$ points (corner of the Brillouin zone of the top bilayer) becomes smaller, while the gap at the $\kappa$ point (corner of the Brillouin zone of the top bilayer) widens.

\newpage
\section{Device}
\begin{figure}[h]
    \centering
    \includegraphics[width=1.0\textwidth]{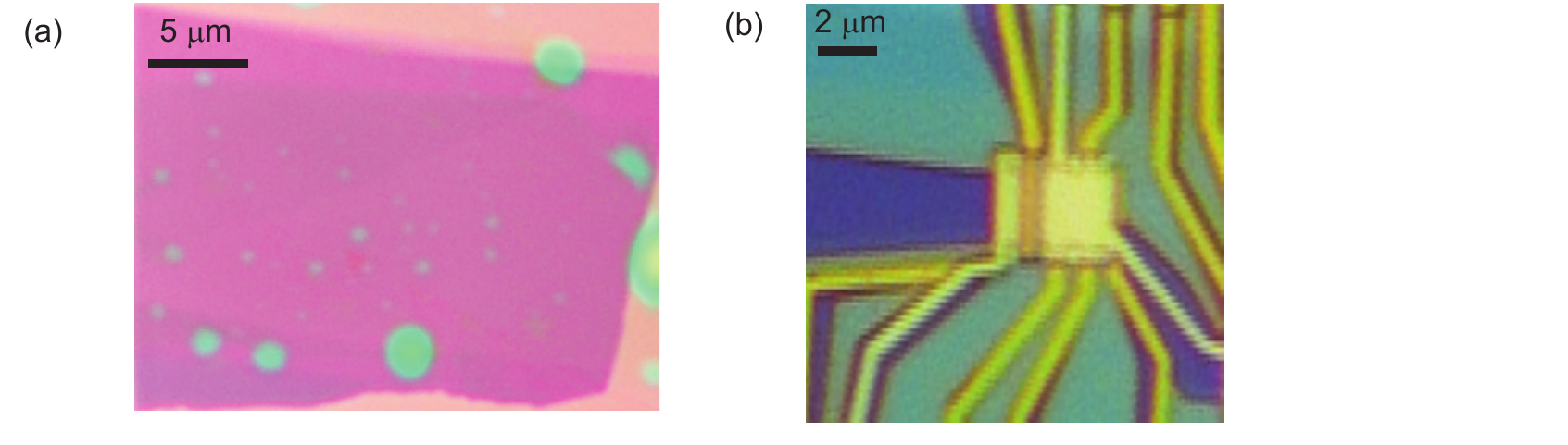}
    \caption{(a) Optical image of the stack. The bubbles are well visible. (b) Optical image of the finished device.
    } 
\label{fig:sup_fig0}
\end{figure}

To prepare the van der Waals heterostack, we use an AFM needle to cut a bilayer graphene flake into two smaller flakes. Then with the dry-transfer method the bilayers are twisted and stacked, encapsulated in hexagonal boron nitride, and, after a graphite layer is added, transferred onto a $\mathrm{Si/SiO}_2$ substrate. After inspection (Fig.~\ref{fig:sup_fig0}), devices are carefully placed in bubble free areas.
Then, the Hall bar mesa is defined with reactive ion etching, after which Ohmic edge contacts are formed using reactive ion etching followed by evaporation of Cr/Au (10/70 nm). The Cr/Au (10/100 nm) top gate is separated from the stack by a dielectric AlO$_\mathrm{x}$ (40 nm) layer.
We used the exact same device as reported on in Ref.~\cite{twistersBZ_2021}.

\clearpage
\newpage

\section{Capacitor model - calculation of $n$ and $D$.}

The capacitor model is an electrostatic description of the multi-layer system where the device stack is assumed to be a series of parallel plate capacitors, influenced by the voltages applied to top ($\Vtg$) and bottom ($\Vbg$) gate.
The TDBG stack can be modeled as a single charge sheet coupled to the top and bottom gate with respective capacitances, $\Ctg$ and $\Cbg$. The total density in the stack and displacement field can be calculated with the following equations:
\begin{equation}
n = \frac{1}{e} (\Cbg \Vbg + \Ctg \Vtg) \\
\end{equation}
\begin{equation}
D = \frac{1}{2\epsilon_0} (\Cbg \Vbg - \Ctg \Vtg)
\label{eqn:d_field}
\end{equation}
A more detailed model can be used in the configuration where wavefunctions are layer polarized and both bilayers are occupied. Then, each bilayer is modeled with a quantum capacitance $C_\mathrm{q} \propto m_{\mathrm{eff}}$ where $m_{\mathrm{eff}} = 0.06 m_e$. Their mutual electrostatic influence is modeled with a capacitance $C_\mathrm{m}$ and the influence of top and bottom gate to the respective layers is modeled with $\Ctg$ and $\Cbg$. The result of the model are densities in both bilayers as function of gate voltages:
\begin{equation}
n_\mathrm{t}=a_\mathrm{t} V_{bg}+b_\mathrm{t} V_{tg}
\label{eqn:n_top}
\end{equation}
\begin{equation}
n_\mathrm{b}=a_\mathrm{b} V_{bg}+b_\mathrm{b} V_{tg}
\label{eqn:n_bottom}
\end{equation}
where $a_\mathrm{t}$, $a_\mathrm{b}$, $b_\mathrm{t}$ and $b_\mathrm{b}$ are functions of $C_q$, $C_m$, $\Ctg$ and $\Cbg$.
Details of the model can be found in \cite{rickhaus2020electronic}.

\section{Screening model}
To show that the hexagonal pattern observed in the in the Fig.~1(d) is a result of screening between the bilayers we add screening to our capacitance model. As a result of the capacitance model, we can relate the densities in the top and bottom bilayer, $n_\mathrm{t}$ and $n_\mathrm{b}$, with the applied top and bottom gate voltages, $\Vtg$ and $\Vbg$, with the following equations \cite{rickhaus2020electronic}:
\begin{equation}\begin{aligned}
e V_{t g} &=\frac{e^{2} n_\mathrm{t}}{\Ctg}+\frac{C_\mathrm{m}}{\Ctg}\left[E_\mathrm{F}\left(n_\mathrm{t}\right)-E_\mathrm{F}\left(n_\mathrm{b}\right)\right]+E_\mathrm{F}\left(n_\mathrm{t}\right)
\label{eqn:el_stat_top}
\end{aligned}\end{equation}
\begin{equation}\begin{aligned}
e \Vbg &=\frac{e^{2} n_\mathrm{b}}{\Cbg}-\frac{C_\mathrm{m}}{\Cbg}\left[E_\mathrm{F}\left(n_\mathrm{t}\right)-E_\mathrm{F}\left(n_\mathrm{b}\right)\right]+E_\mathrm{F}\left(n_\mathrm{b}\right)
\label{eqn:el_stat_bottom}
\end{aligned}\end{equation}
where $C_\mathrm{m}$ is the capacitance between the top and bottom bilayer, $\Ctg$ ($\Cbg$) is the capacitance between the top (bottom) gate and the top (bottom) bilayer, and $E_\mathrm{F} (n_\mathrm{t})$ ($E_\mathrm{F} (n_\mathrm{b})$) is the Fermi energy in the top (bottom) bilayer. The relation between the densities and the Fermi energy is determined by the thermodynamic density of states \cite{ihn2010semiconductor}:
\begin{equation}
\frac{d n_{\mathrm{t,b}}}{d E_{\mathrm{F}}} = \frac{2 m_{\mathrm{eff}}}{\pi \hbar^2} \left[1+2 \sum_{s=1}^{\infty}(-1)^{s} e^{-\pi s / (\omega_{\mathrm{c}} \tau_{q})} \frac{X_{s}}{\sinh X_{s}} \cos \left(\frac{2 \pi s E_{\mathrm{F}}}{\hbar \omega_{\mathrm{c}}}\right)\right]
\label{eqn:SdH_dos}
\end{equation}
where $X_s = 2 \pi^{2} s k_{\mathrm{B}} T / \hbar \omega_{\mathrm{c}}$. Here we assumed a parabolic dispersion with $m_{\mathrm{eff}} = 0.06 m_e$ which we estimated from our MISO model fit.

\begin{figure}
    \centering
    \includegraphics{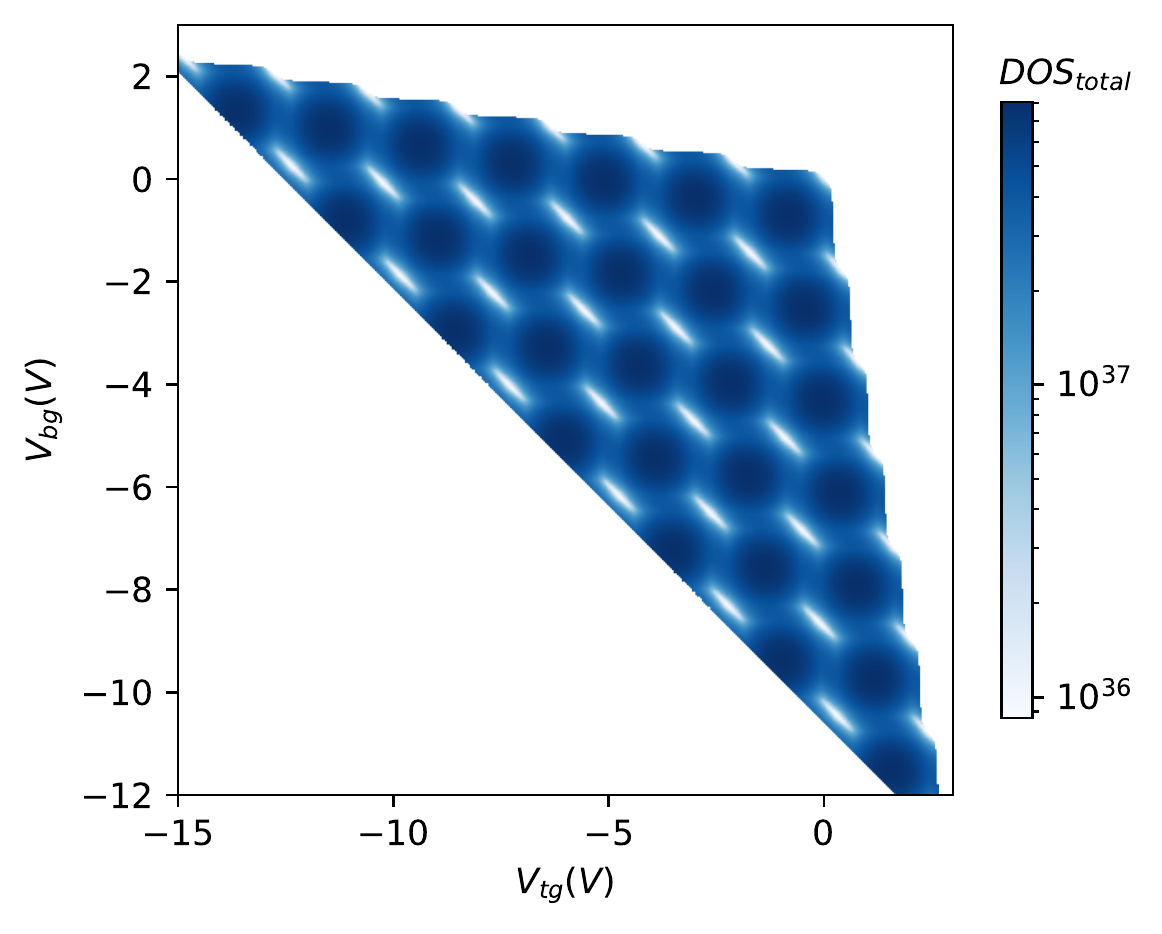}
    \caption{Plot of total density of stats as numerical solution to equations \ref{eqn:el_stat_top} and \ref{eqn:el_stat_bottom}. The calculation was performed on a 1001 by 1001 grid of top and bottom densities with the range of 0 to \num{-3e12} $cm^{-2}$. The parameters used in calculation: $B=\SI{2}{T}$, $m_{\mathrm{eff}} = 0.06 m_e$, $\tau_q = \SI{4e-12}{s}$, $T=\SI{1.2}{K}$ and $s_{\mathrm{max}} = 50$. The plot is truncated at Lifshitz transition, corresponding to density $n=\SI{2.8e12}{cm^{-2}}$ }
\label{fig:sup_fig_hex}
\end{figure}

Since total density of states is reflected in the transport, and hence $R_{xx}$, the goal of the calculation is to relate the total density of states to $\Vtg$ and $\Vbg$. The calculation starts on two dimensional density matrix ($n_\mathrm{t}$, $n_\mathrm{b}$) with the range of densities from $n_{\mathrm{min}}=0$ to $n_{\mathrm{max}}=\SI{-3e12}{cm^{-2}}$ which is approximately the density at the Lifshitz transition. We then numerically calculate the thermodynamic density of states ($DoS (n_\mathrm{t})$ , $DoS (n_\mathrm{b})$) and the corresponding Fermi energies ($E_\mathrm{F} (n_\mathrm{t})$ , $E_\mathrm{F} (n_\mathrm{b})$) for the density matrix. Now, we can use the equations \ref{eqn:el_stat_top} and \ref{eqn:el_stat_bottom} to calculate the respective $\Vtg$ and $\Vbg$ matrices. In the last step, we interpolate the total density of states, which is the sum of density of states in the top and bottom bilayer, on the grid of top and bottom gate voltages. An example of this calculation is shown on Fig.~\ref{fig:sup_fig_hex}, which indicates that screening between the bilayers gives rise to a hexagonal pattern which we observed in the Fig. 1(d) in the main text.

\clearpage
\newpage
\section{Background subtraction}
We subtract the smooth background in measured resistance to reveal the MISO. The background is determined using the Savitsky Golay filter \cite{savitzky1964smoothing}. For Fig. 1(e) and Fig. 1(f) in the main text the parameters of the filter are constant, 33 points and a $3^{rd}$ degree polynomial for both the $\Vtg$ and $\Vbg$ axis. Then, background is determined as an average of filtered data in both the $\Vtg$ and $\Vbg$ axis. Finally, the amplitude of MISO is obtained by subtracting the background from measured data.

For Fig. 2(b) in the main text the parameters of the filter are varying with magnetic field because the fundamental frequency of oscillations is changing. The polynomial degree of 3 is constant, while the point range of the filter is linearly increasing from 7 at $B=\SI{0}{T}$ to 91 at $B=\SI{5}{T}$. The background is defined as the filtered data in $D$ axis (no filtering in $B$ axis) and the amplitude as the difference between the measured data and the background.

Finally, after determining the amplitude of the signal, the data is transformed from the $\Vtg$ and $\Vbg$ axis to the $D$ and $n$ axis using our capacitance model. 
\begin{figure}[h!]
    \centering
    \includegraphics[width=1.0\textwidth]{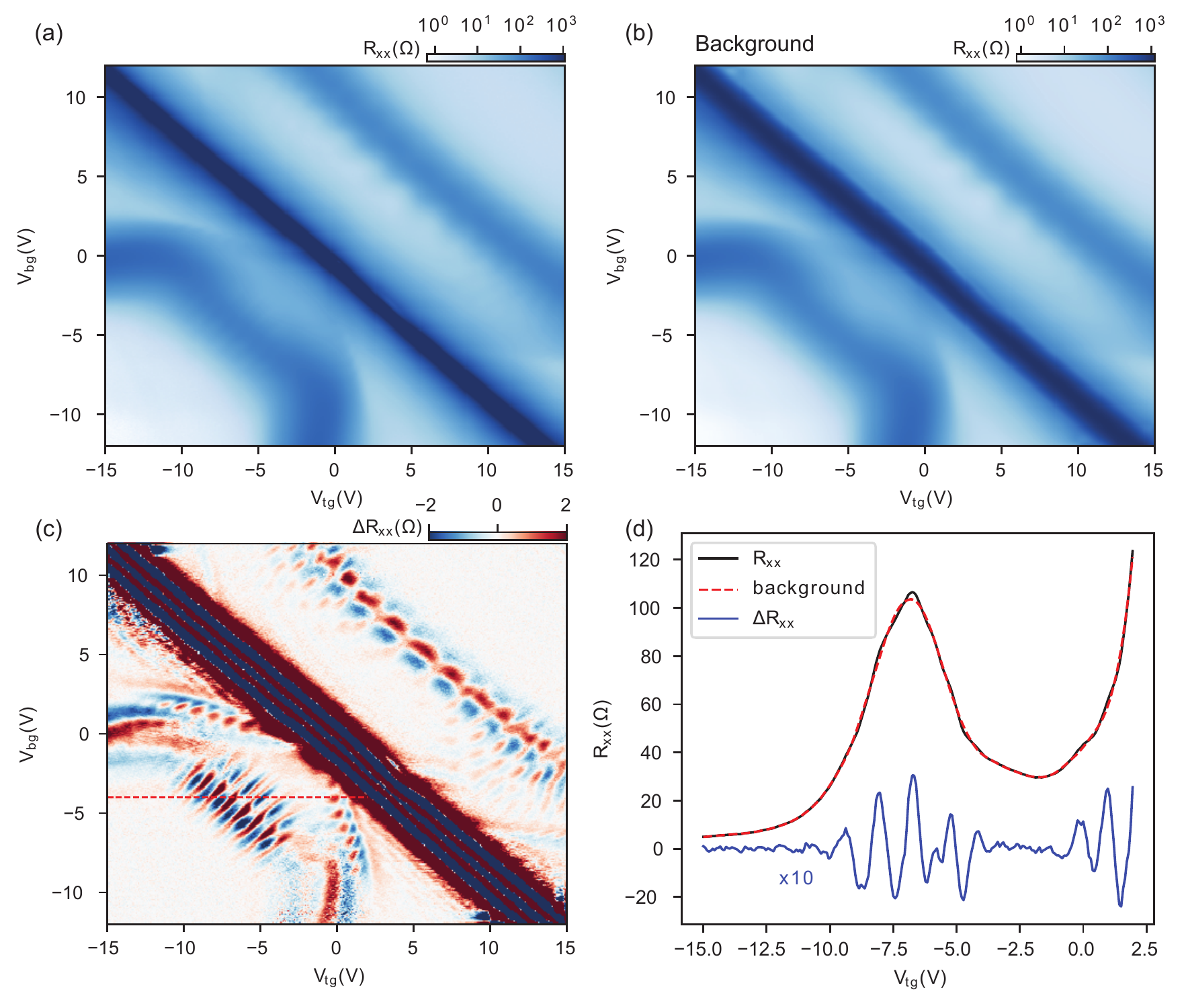}
    \caption{(a) Resistance $R_{xx}$ measured as a function of $\Vtg$ and $\Vbg$ at constant magnetic field $B=\SI{2}{T}$ and temperature $T=\SI{10}{K}$.
    (b) Background $R_{xx}$ obtained by filtering the data in (a).
    (c) Resistance amplitude $\Delta R_{xx}$ obtained by subtracting (b) from (a).
    (d) Example of measured signal $R_{xx}$, obtained background and the resistance amplitude $\Delta R_{xx}$ for $\Vbg=\SI{-4}{V}$, denoted with red dashed line in (c).
    }
\label{fig:sup_fig1}
\end{figure}
\begin{figure}[h!]
    \centering
    \includegraphics[width=1.0\textwidth]{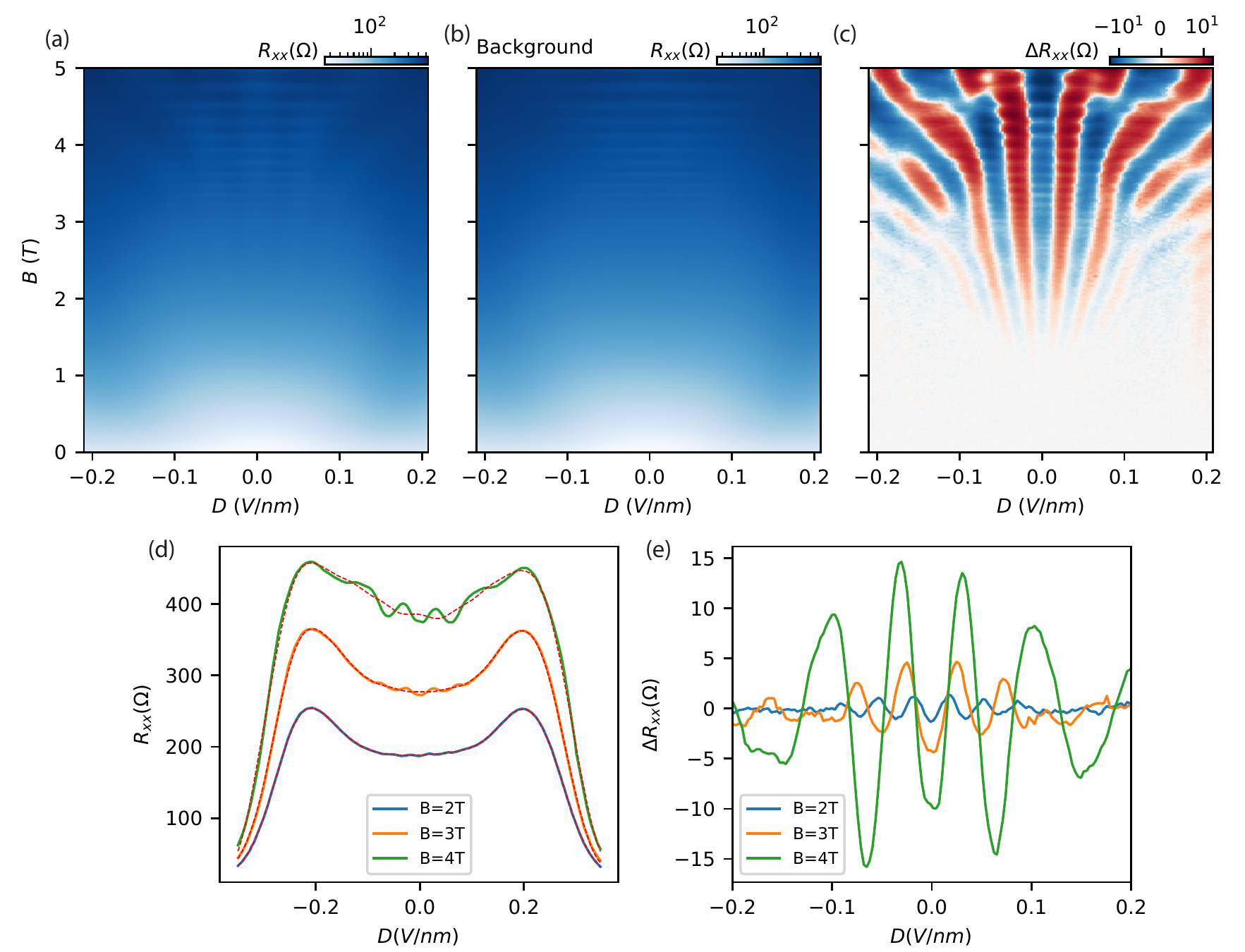}
    \caption{(a) Resistance $R_{xx}$ measured as a function of $B$ and $D$ at constant density $n=\SI{-2.25e12}{cm^{-2}}$ and temperature $T=\SI{20}{K}$.
    (b) Background $R_{xx}$ determined by filtering the data in (a).
    (c) Resistance amplitude $\Delta R_{xx}$ obtained by subtracting (b) from (a).
    (d) Example of measured signal $R_{xx}$ and respective background for magnetic fields of 2, 3 and 4T from data in (a) and (b).
    (e) Resistance amplitude $\Delta R_{xx}$ corresponding to data in (d).
    } 
\label{fig:sup_fig3}
\end{figure}

\clearpage
\newpage
\section{MISO model}
This basic MISO model represents the alignment condition $\Delta (D) = k \hbar e B/m_\mathrm{eff}$. The condition specifies values of $D$ and $B$ for which Landau levels in both minivalleys are aligned i.e. when energy offset $\Delta (D)$ between the minivalleys, which depends on the displacement field, is equal to the integer multiple of the Landau level separation $\hbar \omega_c$.

To relate the displacement field to the $\Delta(D)$ we start by assuming a parabolic dispersion in both bilayers determined by $m_{eff} = 0.06 m_e$. This allows us to write the following equations:
\begin{equation}
(E_{F}-E_{\kappa}) \mathrm { DoS }=n_{\kappa}
\label{eqn:miso_top}
\end{equation}
\begin{equation}
(E_{F}-E_{\kappa'}) \mathrm { DoS }=n_{\kappa'}
\label{eqn:miso_bottom}
\end{equation}
\begin{equation}
\Delta=E_{\kappa'}-E_{\kappa}=\frac{\hbar^{2} \pi}{2 m_{\mathrm{eff}}}(n_{\kappa}(D) - n_{\kappa'}(D))
\label{eqn:miso_model}
\end{equation}
where $E_\mathrm{F}$ is the Fermi energy, $E_{\kappa}$ and $E_{\kappa'}$ are energy maxima of $\kappa$ and $\kappa'$ valence band minivalleys with their respective densities $n_{\kappa}$ and $n_{\kappa'}$. Densities $n_{\kappa}(D)$ and $n_{\kappa'}(D)$ can be calculated using the capacitor model described above using equations \ref{eqn:n_top}, \ref{eqn:n_bottom} and \ref{eqn:d_field} since in this regime $n_{\kappa} = n_{t}$ and $n_{\kappa'} = n_{b}$.
Therefore, equation \ref{eqn:miso_model} allows us to fully determine the displacement field dependence of $\Delta$.

The model fits well for moderate absolute values of $B$ and $D$. At higher $B$, SdHO grow in amplitude and become the dominant magnetotransport phenomenon, while at higher $D$, we suspect, a mismatch in the effective mass between the minivalleys leads to breaking of the MISO alignment condition.
\begin{figure}[h!]
    \centering
    \includegraphics{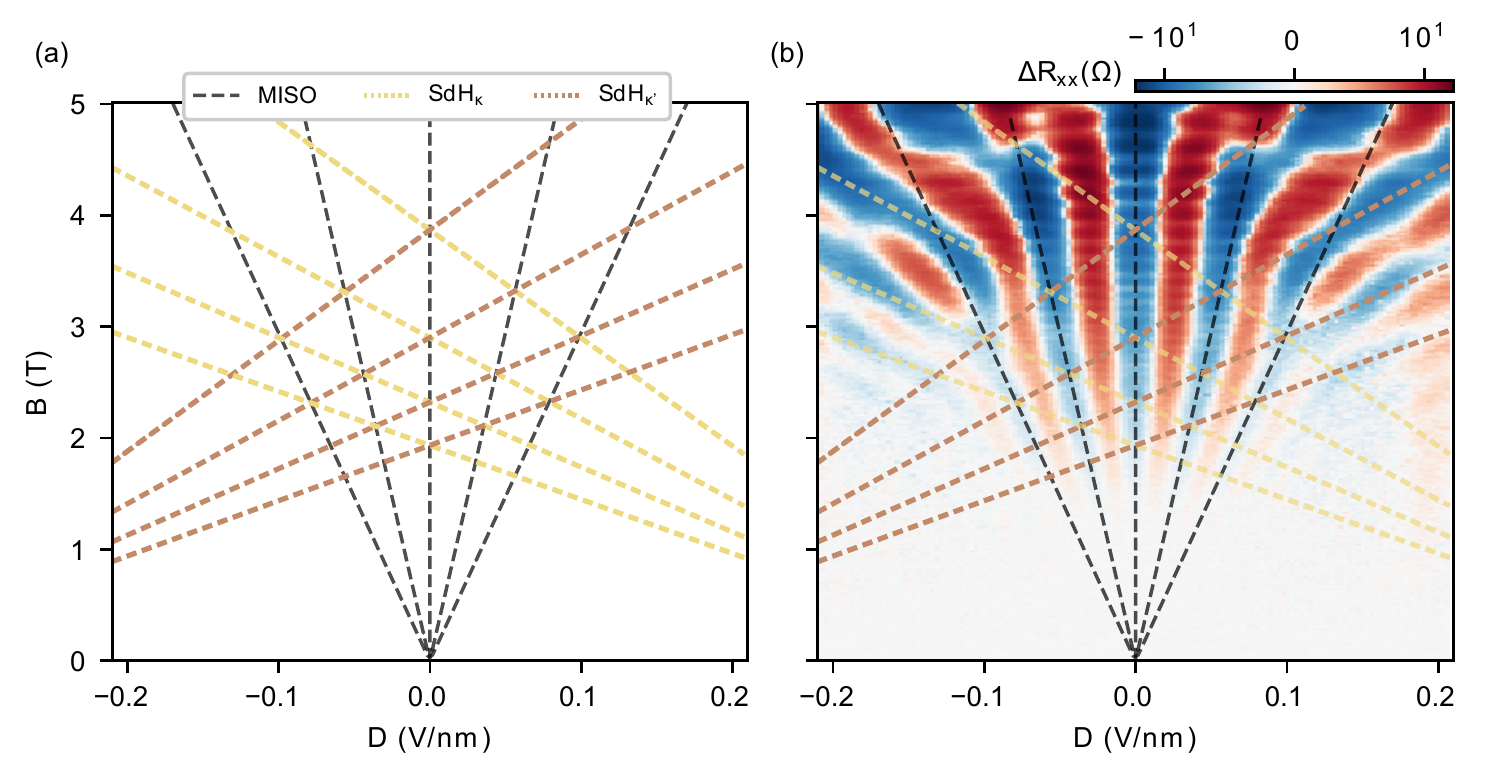}
    \caption{(a) Black dashed lines represent the simple MISO model while yellow (bronze) dashed lines represent SdH oscillations related to $\kappa$ ($\kappa'$) minivalley respectively. 
    (b) Model from (a) plotted on top of the data from \ref{fig:sup_fig3}(c).} 
\label{fig:sup_fig4}
\end{figure}

\newpage
\section{Temperature dependence of resistivity}
\begin{figure}
\begin{center}
\includegraphics[width=0.8\columnwidth]{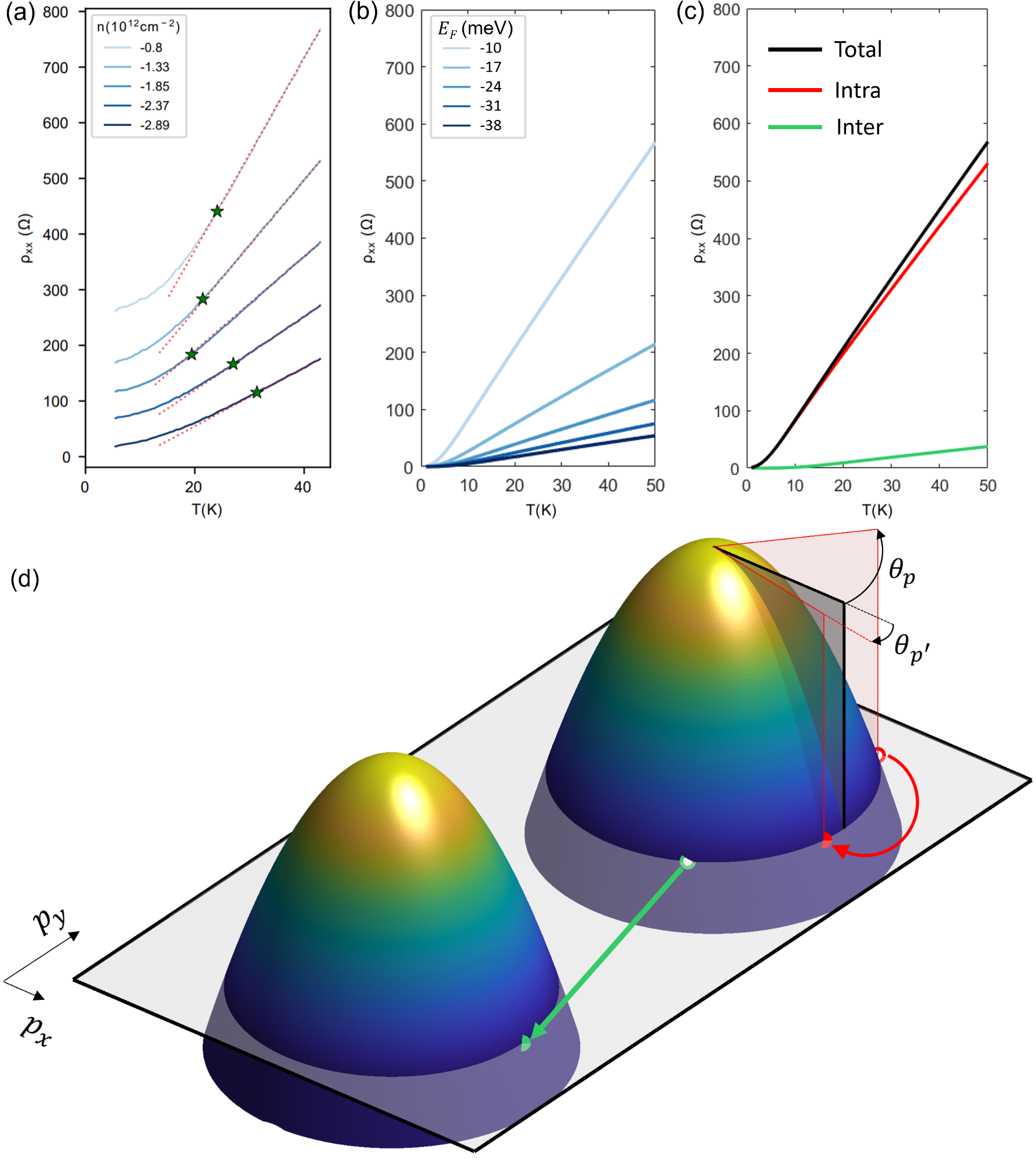}
\caption{(a) Measured longitudinal resistivity $\rho_{xx}$ as a function of temperature $T$ for different densities $n$ as indicated, offset between each other by $50 \Omega$. The red dotted line represents a linear fit, and the green star the onset temperature of the linear slope. (b) Temperature dependence of resistivity simulation using Eq. (\ref{eq:resistivity}), taking $v=8.5\cdot10^{5}$ m/s and $\gamma_1=390$ meV. The five values for the Fermi level correspond to the five doping levels tested in the experiment. (c) Total resistivity for $E_F=10$ meV (black), contribution to the resistivity from scattering processes inside a Bernal bilayer (red), and contribution to the resistivity coming from scattering processes across the twisted interface (green). (d) Band structure of twisted doble bilayer graphene, according to the first term in Eq. (\ref{eq:Ham}), highlighting the two possible scattering processes, intra-bilayer (red) and inter-bilayer (green).
\label{fig:resistivity}}
\end{center}
\end{figure}

We investigate the main scattering mechanisms in region A by measuring the temperature dependence of $\rho_{xx}(T)$ for $D=0$ and $B=0$ with density as parameter. In the Fig.~\ref{fig:resistivity}(a) we plot $\rho_{xx}$ as a function of temperature for different values of density $n$ offset between each other by $50 \Omega$. This data is corrected for a kink around $T=\SI{36}{K}$ caused by the pressure instabilities in the He pot. The correction is performed using a calibration measurement at a single density.

We calculated $\rho_{xx}$ taking the width of the mesa $W=\SI{2374}{nm}$, and the average distance between the contacts of $L=\SI{910}{nm}$. Due to this imperfect geometry the value of $\rho_{xx}$ could be off. Since it does not change the order of magnitude, it will not influence the conclusion below. Green stars indicate the onset of the linear trend (highlighted with red dashed lines) which we interpret as Bloch–Grüneisen temperature \cite{bloch1930elektrischen, gruneisen1933abhangigkeit}. A linear temperature dependence suggests the dominant scattering mechanism is of the electron-phonon type.

We model resistivity in twisted double bilayer graphene assuming that the scattering between electrons and in-plane phonons are the main source for resistivity distinguishing inter- and intra- bilayer scattering mechanisms. The simulation results are presented in Fig.~\ref{fig:resistivity}(b,c). We observe that the model qualitatively predicts the linear dependence at large temperatures, and captures the trend of decreasing resistivity for larger doping levels. Additionally, our model predicts that intra bilayer scattering processes contribute the most to the resistivity [see Fig.~\ref{fig:resistivity}(c)]. This is due to the fact that the coupling constant across the twisted interface is smaller and the momentum of the phonons are typically larger, as shown in Fig.~\ref{fig:resistivity}(d), and their thermal weight is smaller. However, the simulated resistivity is consistently lower than the experimentally observed. Additionally, the onset of linear dependence predicted by the model is significantly lower. This suggests that other sources of resistivity, for example the electron coupling with flexural phonons, could be responsible for this.

\subsection{Resistivity model}
Twisted double bilayer graphene (tDBLG) consists of two Bernal-aligned graphene bilayers stacked on top of each other with their rotational axes tilted by and angle $\theta$. We use plane-wave representation of Bloch functions in the layer $l=1,\dots,4$ with sublattice $\lambda=A,B$ as \cite{garcia2021}
\begin{align}\label{eq:Bloch_PW}
\Psi_{\lambda,\boldsymbol{k}}^{l}(\boldsymbol{r},z) \approx &\frac{e^{i\boldsymbol{k}\cdot\boldsymbol{r}}}{\sqrt{N}\Omega} \sum^2_{j=0} e^{i\boldsymbol{K}^{(j)}_{\mathrm{t/b},\xi}\cdot(\boldsymbol{r}-
\boldsymbol{\tau}_\lambda^l)} 
\mathcal{F}^l (\boldsymbol{K}^{(j)}_{\mathrm{t/b},\xi}+\boldsymbol{k}).
\end{align}
Above, $N$ is the number of unit cells, $\Omega$ the area of the unit cell, the corners of the Brillouin zone for the top/bottom bilayers, $\boldsymbol{K}_{\mathrm{t/b},\xi}^{(j)}= \xi 4\pi/3a [\cos(2\pi j/3), -\sin(2\pi j/3)]_{\pm\theta/2}$ ($a$ is the lattice constant and $j=0,1,2$), are rotated $\pm\theta/2$, and $\mathcal{F} ^l(\boldsymbol{Q},z)$ is the 2D Fourier of the $p_z$ orbital in the layer $l$ evaluated at $\boldsymbol{Q}$\cite{garcia2021}. The atomic position of the two sublattices are $\boldsymbol{\tau}^{1,3}_B=\boldsymbol{\tau}^{2,4}_A=0$ for the dimer sites and $\boldsymbol{\tau}^{1,3}_A=-\boldsymbol{\tau}^{2,4}_B=(0,1)a/\sqrt(3)$ for the non-dimer sites.

We account for electron-phonon interaction by introducing a small displacement vector in the position of an atom in the layer $l$, $\boldsymbol{r}\to\boldsymbol{r}+\boldsymbol{u}^l$, produced by the phonon field
\begin{align}\label{eq:phonon_field}
\boldsymbol{u}^l(\boldsymbol{r},t)&=
\sum_{\boldsymbol{q},\Lambda}
\sqrt{\frac{\hbar^2}{\rho_g\Omega N\omega_{\boldsymbol{q}}}}
(\boldsymbol{\Lambda}
a_{\boldsymbol{q},\Lambda}^le^{-\frac{i}{\hbar}\omega_{\boldsymbol{q},\Lambda}t}+
\boldsymbol{\Lambda}
a^{l\,\dagger}_{-\boldsymbol{q},\Lambda}e^{\frac{i}{\hbar}\omega_{\boldsymbol{q},\Lambda}t})
e^{\frac{i}{\hbar}\boldsymbol{q}\cdot\boldsymbol{r}},
\end{align}
where $\rho_g$ is the density of graphene, $a_{\boldsymbol{q,\Lambda}}^l$ ($a_{\boldsymbol{q,\Lambda}}^{l\,\dagger}$) is the annihilation (creation) operator of a phonon mode in the layer $l$, with momentum $\boldsymbol{q}$, polarisation vector $\boldsymbol{\Lambda}$, which can be longitudinal ($\boldsymbol{\Lambda}=\boldsymbol{q}/q$) or transversal ($\boldsymbol{\Lambda}=\hat{\boldsymbol{z}}\times\boldsymbol{q}/q$), and it has energy $\omega_{\boldsymbol{q}}=s^\Lambda q$, with $s^T=0.09 \,\mathrm{eV\AA}$ and $s^L=0.14 \,\mathrm{eV\AA}$ being the speed of sound for transversal (longitudinal) phonons \cite{cong2019}.

We construct the matrix elements of the Hamiltonian of tDBLG and electron-phonon interaction following \cite{garcia2021}, and adopt an effective low-energy description for electronic states in the Bernal aligned bilayers, neglecting the moire coupling,
\begin{align}\label{eq:Ham}
\mathcal{H}\approx&
\left(
\begin{matrix}
0&\frac{v^2{\pi_{\mathrm{t},\xi}^*}^2}{\gamma_1}&0&0\\
\frac{v^2{\pi_{\mathrm{t},\xi}}^2}{\gamma_1}&0&0&0\\
0&0&
0&\frac{v^2{\pi_{\mathrm{b},\xi}^*}^2}{\gamma_1}\\
0&0&\frac{v^2{\pi_{\mathrm{b},\xi}}^2}{\gamma_1}&0
\end{matrix}
\right)+
\left(
\begin{matrix}
0&\mathrm{V}^{\Lambda}_{3,4}&0&0\\
\mathrm{V}^{\Lambda\,*}_{3,4}&0&\mathcal{V}^{\Lambda}_\theta&0\\
0&\mathcal{V}^{\Lambda\,*}_\theta&0&\mathrm{V}^{\Lambda}_{1,2}\\
0&0&\mathrm{V}^{\Lambda\,*}_{1,2}&0
\end{matrix}
\right),\\
\mathrm{V}^{\Lambda}_{n,m}=&
\sum_{l=n,m}
\frac{\gamma_1K}{2}
\left(i\xi u_x^l-u_y^l\right),\qquad
\mathcal{V}_{\theta}^{\Lambda}=
\frac{\gamma_1}{3}
\sum_{l=2}^3
\sum_{j=0}^2
i\boldsymbol{K}_\xi^{(j)}
\cdot
\boldsymbol{u}^l\,
e^{i\xi\frac{2\pi}{3}j}
e^{i(\Delta\boldsymbol{K}_\xi^{(j)}
-\Delta\boldsymbol{K}_\xi^{(0)})\cdot \boldsymbol{r}}.\nonumber
\end{align}
Here, $\Delta\boldsymbol{K}_\xi^{(j)}\equiv\boldsymbol{K}_{\mathrm{t},\xi}^{(j)}-\boldsymbol{K}_{\mathrm{b},\xi}^{(j)}$ is the wavenumber mismatch between the Bernal aligned bilayers.

Following previous works \cite{wallbank2019}, we model the resistivity examining the flow of electrons under an external electric field applied, $\boldsymbol{E}$, using the Boltzmann equation, 
\begin{align}\label{eq:Boltzmann}
\frac{\mathrm{d}f(\boldsymbol{r},\boldsymbol{p},t)}{\mathrm{d}t}=
\left(
\frac{\partial}{\partial t}+
\boldsymbol{v}
\cdot
\boldsymbol{\nabla}_{\boldsymbol{r}}-
e\boldsymbol{E}
\cdot
\boldsymbol{\nabla}_{\boldsymbol{p}}
\right)f
(\boldsymbol{r},\boldsymbol{p},t)=
I
\left\{
f(\boldsymbol{r},\boldsymbol{p},t)
\right\},
\end{align}
where $f(\boldsymbol{r},\boldsymbol{p},t)$ is the out-of-equilibrium distribution function for electrons located at $\boldsymbol{r}$, with momentum $\boldsymbol{p}$ at time $t$, and $e$ is the elementary charge. The collision integral, $I
\left\{
f(\boldsymbol{r},\boldsymbol{p},t)
\right\}$, is the rate at which the electron occupation number increases or decreases due to scattering events. In this work, we only consider inelastic scattering of electrons with in-plane phonons. If the external electric field is weak enough, the electronic distribution can be written as
\begin{align}\label{eq:f}
f(\boldsymbol{r},\boldsymbol{p},t)\approx
f^{0}(\epsilon_{\boldsymbol{p}}-\varphi_{\boldsymbol{p}})\approx
f^{0}(\epsilon_{\boldsymbol{p}})-
\frac{\partial f^0(\epsilon_{\boldsymbol{p}})}
{\partial \epsilon_{\boldsymbol{p}}}
\varphi_{\boldsymbol{p}}.
\end{align}
The first term in the right hand side is the Fermi-Dirac distribution at the energy $\epsilon_{\boldsymbol{p}}$, while the second term captures the response of the system to the in-plane electric field applied. Because the collision integral of the Fermi distribution is zero, we can write
\begin{align}
I\left\{
f(\boldsymbol{r},\boldsymbol{p},t)
\right\}\approx
I\left\{
-\frac{\partial f^0(\epsilon_{\boldsymbol{p}})}
{\partial \epsilon_{\boldsymbol{p}}}
\varphi_{\boldsymbol{p}}
\right\}\equiv
I_{\boldsymbol{p}}.
\end{align}
 Inserting Eq. (\ref{eq:f}) into Eq. (\ref{eq:Boltzmann}) gives
\begin{align}\label{eq:Boltzmann_lin}
-e
\boldsymbol{E}
\cdot
\boldsymbol{v}_F({\boldsymbol{p}})
\frac{\partial f^0(\epsilon_{\boldsymbol{p}})}
{\partial \epsilon}=
I_{\boldsymbol{p}},
\end{align}
where $\boldsymbol{v}_F(\boldsymbol{p})$ is the Fermi velocity at $\boldsymbol{p}$. To construct the collision integral we integrate over all scattering processes that increase or decrease the electronic distribution with momentum $\boldsymbol{p}$ \cite{landau1980,ziman1960},
\begin{align}
&I_{\boldsymbol{p}}
\approx
\frac{\pi\hbar}{\rho_g}
\sum_{\eta,\Lambda}
\int
\frac{\mathrm{d}{\boldsymbol{p}'}}{(2\pi\hbar)^2}
\left(
2\left|
\hat{W}^{\Lambda,\mathrm{intra}}_{\boldsymbol{p}'\leftarrow\boldsymbol{p}}
\right|^2+
2\left|
\hat{W}^{\Lambda,\mathrm{inter}}_{\boldsymbol{p}'\leftarrow\boldsymbol{p}}
\right|^2
\right)\\
&\qquad\qquad\times
\frac{\partial N^0(\omega_{\boldsymbol{q}})}{\partial\omega_{\boldsymbol{q}}}
\frac{\partial f^0(\epsilon_{\boldsymbol{p}})}
{\partial\epsilon_{\boldsymbol{p}}}
\left(
\varphi_{\boldsymbol{p}'}-\varphi_{\boldsymbol{p}}
\right)
\delta(\epsilon_{\boldsymbol{p}}-\epsilon_{\boldsymbol{p}'}+\eta\omega_{\boldsymbol{q}}^\Lambda),\nonumber\\
&\left|
\hat{W}^{L,\mathrm{intra}}_{\boldsymbol{p}'\leftarrow\boldsymbol{p}}
\right|^2
=
\left[
\frac{\gamma_1 K
\left[-p'
\sin(
2\theta_{\boldsymbol{p}'}+
\theta_{\boldsymbol{p}})+
p\sin
\left(
\theta_{\boldsymbol{p}'}+
2\theta_{\boldsymbol{p}}
\right)
\right]}{2|\boldsymbol{p'-p}|}
\right]^2,\nonumber\\
&\left|
\hat{W}^{T,\mathrm{intra}}_{\boldsymbol{p}'\leftarrow\boldsymbol{p}}
\right|^2
=\left[
\frac{\gamma_1 K
\left[-p'
\cos(
2\theta_{\boldsymbol{p}'}+
\theta_{\boldsymbol{p}})+
p\cos
\left(
\theta_{\boldsymbol{p}'}+
2\theta_{\boldsymbol{p}}
\right)
\right]}{2|\boldsymbol{p'-p}|}
\right]^2,\nonumber
\\
&\left|
\hat{W}^{\Lambda,\mathrm{inter}}_{\boldsymbol{p}'\leftarrow\boldsymbol{p}}
\right|^2
=
\sum_{j=0}^2
\left[
\frac{\gamma_1
\boldsymbol{K}_\xi^{(j)}\cdot
\boldsymbol{\Lambda}}{6}
\right]^2,\nonumber
\end{align}
where $|\boldsymbol{p}\rangle_{\mathrm{t/b}}$ is an eigenvector of top/bottom bilayer, and $N^0(\omega_{\boldsymbol{q}})=
[\exp(\frac{\omega_{\boldsymbol{q}}}{k_BT})-1]^{-1}$ is the occupation number of a phonon with momentum $\boldsymbol{q}$, which is $\boldsymbol{p}-\boldsymbol{p}'$ or $\boldsymbol{p}-\boldsymbol{p}'+\Delta\boldsymbol{K}_\xi^{(j)}$ in intra-bilayer or inter-bilayer scattering processes, respectively. We also introduce $\epsilon_{\boldsymbol{p}}=v^2p^2/\gamma_1$ and $\omega=s^{\Lambda}q$, the electron and phonon energies, respectively, with $s^\Lambda$ being the speed of sound for the phonons. The factor of two in front of the matrix elements arises because each scattering process can generate a phonon two layers. Because we neglected the moire coupling in Eq. (\ref{eq:Ham}), the Fermi surface consists of two circles centred at $(0,\pm\theta K/2)$ [see Fig. \ref{fig:resistivity_T} (d)]. Assuming, without loss of generality, that the electric field is applied along the x direction, $\boldsymbol{E}=(E,0,0)$, the distortion to the Fermi sea is $\varphi_{\boldsymbol{p}}\approx\varphi_p\cos(\theta_{\boldsymbol{p}})$, with $\theta_{\boldsymbol{p}}\arctan(p_y/p_x)$, and Eq. (\ref{eq:Boltzmann_lin}) takes the form
\begin{align}
\frac{E}{\varphi_{p}}=&
\frac{ p}{2\pi\hbar ev_F^2(p)\rho_g}
\sum_{\Lambda}
\int_0^{2\pi}\mathrm{d}\theta_{\boldsymbol{p}'}
\left(
\left|
\hat{W}^{\Lambda,\mathrm{intra}}_{\boldsymbol{p}'\leftarrow\boldsymbol{p}}
\right|^2+
\left|
\hat{W}^{\Lambda,\mathrm{inter}}_{\boldsymbol{p}'\leftarrow\boldsymbol{p}}
\right|^2
\right)
\frac{1-\cos(\theta_{\boldsymbol{p}'}-\theta_{\boldsymbol{p}})}{k_BT\left[\cosh(\frac{\omega_{q}}{k_BT})-1\right]},\nonumber
\end{align}
from which we can compute the resistivity,
\begin{align}\label{eq:resistivity}
\rho^{-1}=&
\frac{8e}{E}
\int\frac{\mathrm{d}\boldsymbol{p}}{(2\pi\hbar)^2}
f(\boldsymbol{r},\boldsymbol{p},t)
\hat{\boldsymbol{x}}\cdot\boldsymbol{v}_F(p)\\
=&
\frac{2e}{(\pi \hbar)^2}
\int_0^{\infty}
p\,\mathrm{d}p
\int_{0}^{2\pi}
\mathrm{d}\theta_{\boldsymbol{p}}\,
\cos^2(\theta_{\boldsymbol{p}})
v_F(p)
\left(-
\frac{\partial f^0(\epsilon_{\boldsymbol{p}})}
{\partial\epsilon_{\boldsymbol{p}}}
\right)
\left[
\frac{E}{\varphi_p}
\right]^{-1}
\nonumber\\
\approx&\frac{16e^2v^2\rho_g}{\pi\hbar\gamma_1}
E_F
\int_{0}^{2\pi}
\mathrm{d}\theta_{\boldsymbol{p}}\,
\cos^2(\theta_{\boldsymbol{p}})\nonumber\\
&\qquad\qquad\times\left\{
\sum_{\Lambda}
\int_0^{2\pi}\mathrm{d}\theta_{\boldsymbol{p}'}
\frac{\left(
\left|
\hat{W}^{\Lambda,\mathrm{intra}}_{\boldsymbol{p}'\leftarrow\boldsymbol{p}}
\right|^2+
\left|
\hat{W}^{\Lambda,\mathrm{inter}}_{\boldsymbol{p}'\leftarrow\boldsymbol{p}}
\right|^2
\right)
\left(1-\cos(\theta_{\boldsymbol{p}'}-\theta_{\boldsymbol{p}})
\right)}{k_BT\left[\cosh(\frac{\omega_{q}}{k_BT})-1\right]}
\right\}^{-1},\nonumber
\end{align}
where the integration is performed around one of the two Fermi pockets, and we take into account the spin, valley and bilayer degeneracy. The resistivity in Eq. (\ref{eq:resistivity}) accounts for all possible scattering processes within the Fermi pocket, determined by the doping $n$.

%